\documentclass[12pt]{memoir}\usepackage[]{graphicx}\usepackage[usenames,dvipsnames]{color}
\makeatletter
\def\maxwidth{ %
  \ifdim\Gin@nat@width>\linewidth
    \linewidth
  \else
    \Gin@nat@width
  \fi
}
\makeatother

\definecolor{fgcolor}{rgb}{0.345, 0.345, 0.345}

\usepackage{framed}
\makeatletter
 {\par\unskip\endMakeFramed%
 \at@end@of@kframe}
\makeatother

\definecolor{shadecolor}{rgb}{.97, .97, .97}
\definecolor{messagecolor}{rgb}{0, 0, 0}
\definecolor{warningcolor}{rgb}{1, 0, 1}
\definecolor{errorcolor}{rgb}{1, 0, 0}
\newenvironment{knitrout}{}{} 

\usepackage{alltt}

\usepackage{cdsc-memoir}
\chapterstyle{cdsc-article}

\usepackage[utf8]{inputenc}
\usepackage{wrapfig}

\usepackage[T1]{fontenc}
\usepackage{textcomp}
\usepackage[garamond]{mathdesign}

\usepackage[letterpaper,left=1.65in,right=1.65in,top=1.3in,bottom=1.2in]{geometry}

\usepackage{graphicx}
\usepackage{enumerate}

\usepackage{dcolumn}
\usepackage{adjustbox}

\usepackage[usenames,dvipsnames]{color}
\usepackage[breaklinks]{hyperref}

\hypersetup{colorlinks=true, linkcolor=Black, citecolor=Black, filecolor=Blue,
    urlcolor=Blue, unicode=true}

\makeatletter
\renewcommand*{\@fnsymbol}[1]{\ensuremath{\ifcase#1\or *\or \dagger\or \ddagger\or
 \mathsection\or \mathparagraph\or \|\or **\or \dagger\dagger
  \or \ddagger\ddagger \else\@ctrerr\fi}}
\makeatother

\usepackage[american]{babel}
\usepackage{csquotes}
\usepackage[natbib=true, style=apa, backend=biber]{biblatex}
\DeclareLanguageMapping{american}{american-apa}

\defbibheading{secbib}[\bibname]{%
  \section*{#1}%
  \markboth{#1}{#1}%
  \baselineskip 14.2pt%
  \prebibhook}

\def\citepos#1{\citeauthor{#1}'s (\citeyear{#1})}

\firmlists

\usepackage{bm}

\gdef\GITAbrHash{60d9c0a}%
\gdef\VCRevision{\GITAbrHash}%
\gdef\VCDateTEX{2021/11/20}%
\IfFileExists{upquote.sty}{\usepackage{upquote}}{}
\begin{document}

\setlength{\parskip}{4.5pt}
\baselineskip 16pt

\title{The Hidden Costs of Requiring Accounts: Quasi-Experimental Evidence from Peer Production}

\author{Benjamin Mako Hill\\
         \href{mailto:makohill@uw.edu}{makohill@uw.edu}\\
         \vspace{0.5em}
         Aaron Shaw\\
         \href{mailto:aaronshaw@northwestern.edu}{aaronshaw@northwestern.edu}
}

\date{}

\published{\textsc{\textcolor{BrickRed}{A version of this manuscript has been published as:}} \textcolor{BrickRed}{
    Hill, Benjamin Mako, and Aaron Shaw. 2020. “The Hidden Costs of Requiring Accounts: Quasi-Experimental Evidence from Peer Production.” \textit{Communication Research}, 48 (6): 771–95.\\
    \href{https://doi.org/10.1177/0093650220910345}{\color{BrickRed}https://doi.org/10.1177/0093650220910345}.}}

\maketitle

\vspace{-2em}

\begin{abstract}
  Online communities like Wikipedia produce valuable public information goods. While some of these communities require would-be contributors to create accounts, many do not. Does this requirement catalyze cooperation or inhibit participation? Prior research provides divergent predictions but little causal evidence. We conduct an empirical test using longitudinal data from 136 natural experiments where would-be contributors to wikis were suddenly required to log in to contribute. Requiring accounts leads to a small increase in account creation, but reduces both high and low quality contributions from registered and unregistered participants. Although the change deters a large portion of low quality participation, the vast majority of deterred contributions are higher quality. We conclude that requiring accounts introduces an under-theorized tradeoff for public goods production in interactive communication systems. \\\vspace{2em}
\end{abstract}

\vspace{-2em}

As a core principle, Wikipedia has always welcomed what it calls ``unregistered editing'' by people who have not created an account on the site.\footnote{\url{https://en.wikipedia.org/wiki/Wikipedia:Welcome_unregistered_editing} (archived at: \url{https://perma.cc/P7EN-2VWH}). 
} Because vandalism and malicious editing persists on Wikipedia---as much as 97\% of it from unregistered editors\footnote{See \url{https://en.wikipedia.org/wiki/Wikipedia:Counter-Vandalism_Unit/Vandalism_studies} (archived at: \url{https://perma.cc/5TEM-6R77}).
The specific estimate depends on the measurement method used.}---Wikipedia participants have repeatedly suggested requiring contributors to create accounts.\footnote{Proposals to ``prohibit anonymous users from editing'' are the first item on Wikipedia's list of perennial proposal topics related to editing: \url{https://en.wikipedia.org/wiki/WP:PEREN} (archived at: \url{https://perma.cc/EZ23-ELJQ}).} Even during a series of crises around defamation, Wikipedia's leaders refused to require Wikipedia contributors to reveal their identity or use consistent pseudonyms \citep{jemielniak_common_2014}.
Wikipedia's commitment to unregistered editing speaks to a longstanding concern in the study of interactive communication systems: do stable identifiers catalyze the creation of public information goods \citep{benkler_wealth_2006, fulk_connective_1996, monge_theories_2003, yuan_individual_2005}?

According to some scholars, stable identifiers support social connections that enable public goods production and prosocial cooperation in exchange networks \citep[e.g., ][]{axelrod_evolution_1984, faraj_network_2011, ostrom_governing_1990, yamagishi_solving_2009}. 
Within online communities, identifiers and reputations increase social transparency, group attachment, trust, accountability, and commitment among members, leading to increased participation \citep{cheshire_social_2008, kraut_building_2012, ren_building_2012}. Even requiring low cost, disposable account names can prevent negative disinhibited behavior, non-cooperation, and strategic gaming of reputation systems in computer-mediated communication without deterring good-faith participants \citep{friedman_social_2001, joinson_causes_1998, kraut_building_2012, walther_computer-mediated_1996}.

Other research argues that minimized contribution costs facilitate large-scale participation \citep{benkler_coases_2002, benkler_wealth_2006, bennett_logic_2012, cheshire_selective_2007, kollock_economies_1999}. In this view, the requirement to create a stable identifier, even a disposable pseudonym, imposes a transaction cost that inhibits cooperation that would otherwise occur.
Unregistered contributions can also introduce valuable perspectives \citep{anthony_reputation_2009, kane_emergent_2014, woolley_evidence_2010} and can stimulate activity from experienced community members \citep{gorbatai_paradox_2014}.

Synthesizing this prior work, we anticipate that requiring peer production contributors to use stable identifiers will decrease both high and low quality contributions. We report the results of an empirical, observational test of these claims. The study exploits a series of 136 natural experiments that occurred in collectives engaged in peer production. Across all of these groups, would-be contributors went from being able to edit without an account to being obligated to create an account and log in to participate. In every case, the feature change was sudden and resulted in the sort of institutional shock that supports the use of quasi-experimental methods to identify causal effects \citep{murnane_methods_2011}. Using exhaustive longitudinal data gathered from public database records, we employ a method inspired by regression discontinuity design (RDD) to estimate the impact of the feature change.

Although requiring accounts causes the proportion of low quality contributions to decrease enormously, the large majority of deterred edits are of higher quality.
These effects hold even among participants who had accounts prior to the intervention, indicating that unregistered contributions stimulate activity among established contributors. Overall, we find no support for the argument that user accounts actively catalyze short term cooperation. 
We conclude that stable identifiers like user accounts introduce previously under-theorized tradeoffs in public information goods production in open collaboration systems.

\subsection{Public Information Goods Production and Stable Identifiers Online}

Networked communication and interaction in online communities has facilitated large scale public information goods production \citep{benkler_coases_2002, fulk_connective_1996, kollock_economies_1999, monge_theories_2003}. Many of the central elements of classical theories of public goods production and commons \citep{olson_logic_1965, ostrom_governing_1990} do not change with the move to online environments and digital resources. However, several key aspects of the communicative practices involved undergo fundamental shifts. Two of the most frequently referenced shifts include (1) radically reduced transaction costs associated with participation \citep[e.g., ][]{benkler_wealth_2006, bennett_logic_2012, fulk_connective_1996} and (2) technological mediation that allows individuals to participate with varying degrees of anonymity \citep[e.g., ][]{anonymous_reveal_1998, joinson_causes_1998, walther_computer-mediated_1996}.\footnote{Although many describe unregistered users of online communities as ``anonymous,'' there are distinct types of anonymity \citep{anonymous_reveal_1998}.} To combat the negative effects of disinhibition due to anonymity, many communities require would-be contributors to create an account and log in (e.g. Reddit). On the other hand, requiring stable identifiers---even disposable pseudonyms---constitutes a cost. Communities like Wikipedia argue that the time required to create an account imposes a prohibitive barrier on would-be contributors.

Our analysis focuses on online collectives engaged in \emph{commons-based peer production}, the form of digitally-networked, open public goods production that has created valuable resources such as Free/Libre Open Source Software, Wikipedia, and OpenStreetMap \citep{benkler_coases_2002, benkler_wealth_2006}. \citet{benkler_peer_2015} define peer production as ``a form of open creation and sharing performed by groups online that: (1) set and execute goals in a decentralized manner; (2) harness a diverse range of participant motivations, particularly non-monetary motivations; and (3) separate governance and management relations from exclusive forms of property and relational contracts.'' Peer production encompasses some of the largest sustained instances of open collaboration
and collective action achieved through Internet-mediated communication \citep{benkler_peer_2015, cheshire_selective_2007, fulk_connective_1996, schweik_internet_2012}. As a result, peer production communities provide ideal environments to analyze the effects of different sociotechnical arrangements on public information good production. 

Existing research makes divergent predictions about the effects of requiring user accounts on public goods production in online communication. One body of prior studies argue that stable identifiers work as catalysts of cooperation by facilitating accountability, group identification, boundaries, and commitment that lead to higher quality participation. Others contend that requiring accounts imposes costly obstacles that deter participation. A related set of studies suggests that unregistered contributions may introduce diverse perspectives and stimulate activity among experienced community members. Thus, reductions in low quality content may lead to overall declines in high quality contribution as well. Our hypotheses, developed below, borrow from all of these approaches and reflect a synthesis of this earlier work.

\subsection{Stable Identifiers as Catalysts of Cooperation}

The value of stable identifiers in public information goods production and interactive communication networks derives from the importance of boundaries for cooperation and common pool resource production \citep{ostrom_governing_1990}. At small scale, stable identifiers facilitate sustained communication in repeated interactions leading to the development of reputations and norms \citep{axelrod_evolution_1984}.
Relying on identifiers, reputation systems can promote accountability and community boundaries in computer-mediated communication. Without identifiers, anonymity online can lead to disinhibition and toxic behavior like spam and hate speech \citep{joinson_causes_1998} which tend to be counter-productive to the creation of large-scale public information goods \citep{kraut_building_2012}. Even light-weight identifiers can facilitate norm enforcement and more supportive interpersonal communication dynamics \citep{anonymous_reveal_1998, walther_computer-mediated_1996}. 

Stable identifiers can also support computer-mediated information sharing by reducing anonymity and enhancing trust and opportunities for social approval \citep{cheshire_selective_2007, kollock_economies_1999}. 
The low costs of reproducing and consuming digital information makes would-be contributors particularly susceptible to social incentives like feedback and status \citep{cheshire_selective_2007, cheshire_social_2008, fulk_connective_1996, kollock_economies_1999}. Reduced anonymity between participants thus enhances the social feedback that sustains cooperation \citep{cheshire_selective_2007, cheshire_social_2008, faraj_network_2011, kollock_economies_1999} and reduces non-cooperative behavior \citep{yamagishi_solving_2009}. Stable identifiers can also enhance identification and commitment online, leading to more robust, sustained participation \citep{kraut_building_2012, ren_building_2012}. Visible reputations support the formation of trust and bonding among participants, who develop a sense of collective identity \citep{ren_building_2012}, reinforcing group boundaries in positive ways.

In digital settings, low quality contributions can be easily undone. However, the process of filtering contributions requires resources and effort that could be allocated to more productive ends. In theory, the requirement that contributors create accounts imposes a barrier analogous to the use of ``CAPTCHA'' insofar as it selectively keeps out vandals and spammers who would not have made productive contributions anyway \citep{kraut_building_2012}. Appropriately ``cheap'' pseudonyms would thus lead to an increase in registered contributors and enhance the overall quality of participation so long as the system cannot be gamed too easily \citep{friedman_social_2001}. Barriers to entry can, in this view, serve informal boundary-maintenance functions, ensuring that would-be contributors have sufficient motivation to overcome small obstacles while individuals unwilling to abide by minimal community norms will not get involved in the first place.

The ``catalysts'' perspective leads us to expect that account requirements should not deter contributions from good-faith participants. In a stronger form, this perspective would predict that contributions from more socially integrated participants would lead to an overall increase in quality.
Based on the approaches described above, a requirement to register an account before contributing to a peer production project should drive away low quality contributions like spam while prompting an increase in the creation of new accounts and either stable or increasing rates of high quality contributions.

\subsection{Barriers to Entry as Transaction Costs and Social Inhibitors}

The speed and ease with which individuals can make small contributions to
digitally-networked communication systems differentiates them from earlier knowledge sharing and management \citep{benkler_wealth_2006, bennett_logic_2012, cheshire_selective_2007, fulk_connective_1996, kollock_economies_1999}. For example, in Wikipedia, it takes a few seconds and very little specialized knowledge to add a missing comma to an article. Cost-oriented explanations of online collaboration argue that such radically low barriers to entry enhance the quantity of participation. Further arguments extend this to propose that even low quality participation stimulates further activity leading to enhanced content quality.

Classical theories of public goods and collective action treat barriers to entry as a resource constraint likely to deter participation \citep{olson_logic_1965, oberschall_social_1973}.
Low participation costs help explain why some collectives achieve critical mass while others do not \citep{marwell_critical_1993}. Reduced participation costs caused by new communication technology also explain the rise of online forms of collective action and group generalized exchange \citep{benkler_wealth_2006, bennett_logic_2012, kollock_economies_1999}. 
In online environments, low participation costs often translate into fluid or open community boundaries. 
Classical theories predict that these open boundaries might facilitate free riding \citep{olson_logic_1965} or undermine community stability and cohesion \citep{ostrom_governing_1990}. However, open communities and crowds pursuing information sharing can benefit from larger group size as well as greater diversity of perspectives \citep{benkler_wealth_2006, bennett_logic_2012, kane_emergent_2014, woolley_evidence_2010, zhang_group_2011}. Open boundaries also make it easier for organizations to build informal, lightweight relationships with volunteers and then leverage opportunities to develop deeper identification and commitment \citep{ren_building_2012}.

The absence of a registration requirement may also facilitate easier onboarding of new participants and greater diversity of contributions. For example, newcomers to online communities often pursue low-commitment, marginal tasks before they create accounts \citep{antin_readers_2010}. The opportunity to contribute to a project without a stable identifier allows participants to experiment without risk or cost to their reputations \citep{anonymous_reveal_1998, walther_computer-mediated_1996}, potentially yielding more innovative and novel contributions \citep{bernstein_4chan_2011}. Unregistered contributions can bring new, diverse knowledge to the table that attracts the attention of more experienced community members \citep{anthony_reputation_2009} and provokes those experienced community members to make further improvements \citep{kane_emergent_2014}. The interactions that result can become self-sustaining \citep{bennett_logic_2012, margolin_emotional_2018} lead to higher quality information goods overall even when unregistered participants add relatively low quality content \citep{gorbatai_paradox_2014}.

Together these approaches contradict the ``catalysts of cooperation'' perspective on stable identifiers. The work emphasizing barriers to entry suggests that a shift to requiring accounts should decrease participation activity in communities, including contributions of both low and high quality. In terms of new account creation, these perspectives do not advance a clear prediction. Finally, research on the productive role of unregistered contributors suggests that requiring identifiers may inhibit overall quality improvements by reducing group size and reducing a source of stimulus for follow-on contributions, leading existing community members to participate less as well.

\subsection{Tradeoffs of Stable Identifiers in Peer Production}

Synthesizing the bodies of research summarized above, we suggest that stable identifiers in the organization of online public information goods entail tradeoffs. On the one hand, stable identifiers can enhance group boundaries, accountability, cohesion, and dynamics of reputation and status. Imposing stable identifiers can drive away spammers, reduce the burden of norm enforcement, and lead to virtuous cycles of engagement. On the other hand, stable identifiers impose a barrier to participation that should reduce the likelihood of contribution among some would-be participants who are not vandals. In losing these would-be participants, communities will also shrink in terms of good faith contributions and members. These reductions in contributions from outsiders, both good and bad, will undermine the activity of established contributors indirectly.

The tradeoffs we anticipate underscore the importance of several questions that prior studies have failed to resolve. On the catalysts side, studies on the beneficial impacts of stable identifiers tend to overlook whether the identifiers deter good faith participation. Meanwhile, studies in the barriers camp have approached transaction costs as a relatively homogeneous concept, failing to make specific predictions about the effects of common types of costs. In the context of peer production and other forms of online collective action, this leads us to question whether requiring accounts for participation would constitute a sufficiently costly barrier to entry that it would drive away otherwise motivated and interested contributors? Alternatively, requiring account creation might, as many project administrators and systems designers believe, keep spammers out while having little or no effect on good faith contributors.

We derive our hypotheses from this tradeoffs perspective. First, drawing from the view that stable identifiers are not sufficiently burdensome to drive away all contributors, we anticipate that \emph{requiring account creation for participation in peer production communities will cause an increase in the number of newly registered accounts} (\emph{H1}).

We also expect that requiring accounts will depress the rate of contributions. Drawing from the theoretical perspective derived above, we divide this into two hypotheses. Based on both the catalysts and costs perspectives, we expect that \emph{requiring account creation for participation in peer production communities will cause a reduction in the number of subsequently removed contributions} (\emph{H2}). Communities require accounts because they believe the effect of doing so will primarily be to provide a costly barrier biased toward deterring low quality and unwanted forms of participation like vandalism and spam. Because vandals and spammers are not motivated through collaboration, their contributions are unlikely to be catalyzed by an account requirement.

Our third hypothesis anticipates that the imposition of the costs associated with creating and signing-in to a user account will also deter would-be contributors who would have added something of value and that \emph{requiring account creation for participation in peer production communities will cause a reduction in high quality contributions} (\emph{H3}). This reflects the view that requiring accounts is a broadly-felt costly barrier whose effect on would-be contributors of high quality knowledge will not be outweighed by the benefits of accounts as catalysts of cooperation.

Finally, building on the argument that unregistered contributors stimulate activity by existing community members by \citet{gorbatai_paradox_2014}, \citet{kane_emergent_2014} and others, we hypothesize that \emph{requiring account creation for participation in peer production communities will cause a reduction in contributions from participants who contributed with accounts prior to the requirement} (\emph{H4}). Directly contradicting the catalysts point of view, this hypothesis also extends the transaction costs argument by proposing an indirect impact of the design change on contributors who already hold accounts.

\section{Methods}

We test these hypotheses by combining observational digital trace data from a sample of 136 peer production projects with an analytic strategy that allows us to identify causal effects. Prior research on the impact of stable identifiers and small transaction costs on public information goods does not compare across multiple communities or incorporate evidence capable of supporting causal claims. Other studies that address causal concerns have incorporated findings from small samples with limited external validity. Additionally, previous work has focused on the effect of transaction costs and stable identifiers on the aggregate \emph{quantity} of participation without differentiating or evaluating its quality \citep{jay_sign_2016}.

The communities in our study underwent a policy change after which contributing without an account became unavailable. We estimate the causal effect of the policy change on communities that previously had unregistered activity by comparing trace data on contributions immediately before and after the change using a quasi-experimental design described below. This approach follows a small number of studies that use changes to communication systems in peer production as a way of drawing causal inference \citep{zhang_group_2011, geiger_when_2013, narayan_all_2019, slivko_identification_2016}.

\subsection{Empirical Setting: Wikis and Wikia Inc.}

Our empirical setting is a large population of peer production communities engaged in the collaborative creation of \emph{wikis}.\footnote{We have previously published research using data from this empirical setting \citep{shaw_laboratories_2014} and some of the the text in this section is adapted from that work.}
The term ``wiki'' refers to a type of software designed to facilitate the collaborative, asynchronous writing and distribution of textual content; the communities that use wiki software; and the products created by these groups \citep{leuf_wiki_2001}. Wikipedia is the most famous example of a wiki, but there are hundreds of thousands of other wikis with different goals, topics, and scopes. Through the enormous success of Wikipedia---one of the five most popular websites in the world---wikis figure prominently in \citepos{benkler_coases_2002} original conceptualization of commons-based peer production and are among the most visible and impactful public information goods online.

We analyze a population of wikis hosted by the for-profit firm \emph{Wikia}. Wikia, started in 2004, sought to apply Wikipedia's production model to a broad range of topics. Wikia was founded by Jimmy Wales, Wikipedia's founder, and Angela Beesley, an active and respected early Wikipedia contributor. Wikia's policies, structures, and technologies have been heavily influenced by Wikipedia. In terms of the content and scope, Wikia wikis vary enormously, addressing popular culture and ``fan culture'' topics as well as things like software, food, and fashion. Some of the largest wikis host information about massive multiplayer online video games like Halo, television shows like Breaking Bad, or cultural phenomena like puppetry. On October 4, 2016, Wikia partially rebranded itself as Fandom. Although many firms host wikis (e.g., PBWiki, WikiSpaces, and SocialText), Wikia is unique in that it only hosts publicly accessible, volunteer-produced content.

Wikia wikis constitute a large, inclusive population of peer production communities \citep{hill_studying_2019}. Wikia relies on peer production to create the content of its websites. All of its wikis invite contributions from the general public and all Wikia content is distributed freely and openly. The vast majority of Wikia communities allow contributions without an account.

\subsection{Data}

In September 2012, Wikia staff shared with us a list of 181 wikis that, according to their records, had changed a configuration setting that required all would-be contributors to create accounts before making an edit. The staff also provided information and documentation that the intervention happened at the request (or at least with the consent) of administrators from the affected communities. They explained that on Wikia, configuration of the wiki was generally left to administrators and that it was the staff's belief that only a small proportion of non-administrators---and no unregistered users---would have been involved in the decision to make the change. We conducted text-based searches on a randomly selected 10\% of wikis in our sample and did not find any evidence that the interventions were announced ahead of time. The fact that the change was implemented in software meant that would-be contributors had no obvious way of avoiding exposure to it.

We collected public datasets on the full contribution histories of each of the affected wikis with the exception of 4 wikis whose database records has been deleted by Wikia. For each wiki, Wikia either publishes the data as an XML database dump or exposes the relevant information to construct an XML dump through their API. The XML dumps contain the full history of every revision made by every user to every page in each wiki.\footnote{The dumps do not include data from pages whose entire history of contribution has been deleted. Since pages consisting only of low quality content can be deleted by administrators in this way, the measures of damage described below should be interpreted as incomplete and conservative.}

The data we received from Wikia staff included only the most recent time that the option for unregistered editing had been updated in the site's configuration database. We found that the dates shared by Wikia staff corresponded to a complete cessation of editing from users without accounts in the vast majority of cases. For 8 wikis, we found compelling evidence that the configuration change had occurred earlier and adjusted the date accordingly. Details of this process, visualizations of the edit histories of these wikis, and additional models in which we do not adjust the dates appear in the ``Adjusted Cut-offs'' section of our online supplement. Making this adjustment does not alter the pattern of our results.

Because our hypotheses concern the effect of requiring accounts on active communities with both registered and unregistered contributors, we restrict our sample to wikis with minimal activity and where at least some contributions were made by unregistered users before they are blocked. We use the following inclusion criteria: wikis must (1) be at least 4 weeks old when the intervention occurred; (2) have received at least 1 contribution from an anonymous contributor in the 8-week period before the change; (3) have at least one contribution in at least 70\% of the weeks during the 16-week period before and after the change; and (4) experience at least a 90\% decrease in the number of contributions from unregistered users in 8-week periods before and after the change indicating that the block actually occurred. We exclude at least one wiki from the main analysis for each of these reasons leaving us with a total of 136 wikis. We provide details on our sample selection, descriptive statistics for the excluded wikis, and models fit on a dataset that does not exclude any of these wikis in the ``Sample Selection'' section of our supplement. Although returning excluded wikis to our analysis reduces the size of our effects slightly, it does not change the substance of our results.

\subsection{Measures}
\label{sec:measures}

\begin{figure}
\begin{knitrout}
\definecolor{shadecolor}{rgb}{0.969, 0.969, 0.969}\color{fgcolor}
\includegraphics[width=\maxwidth]{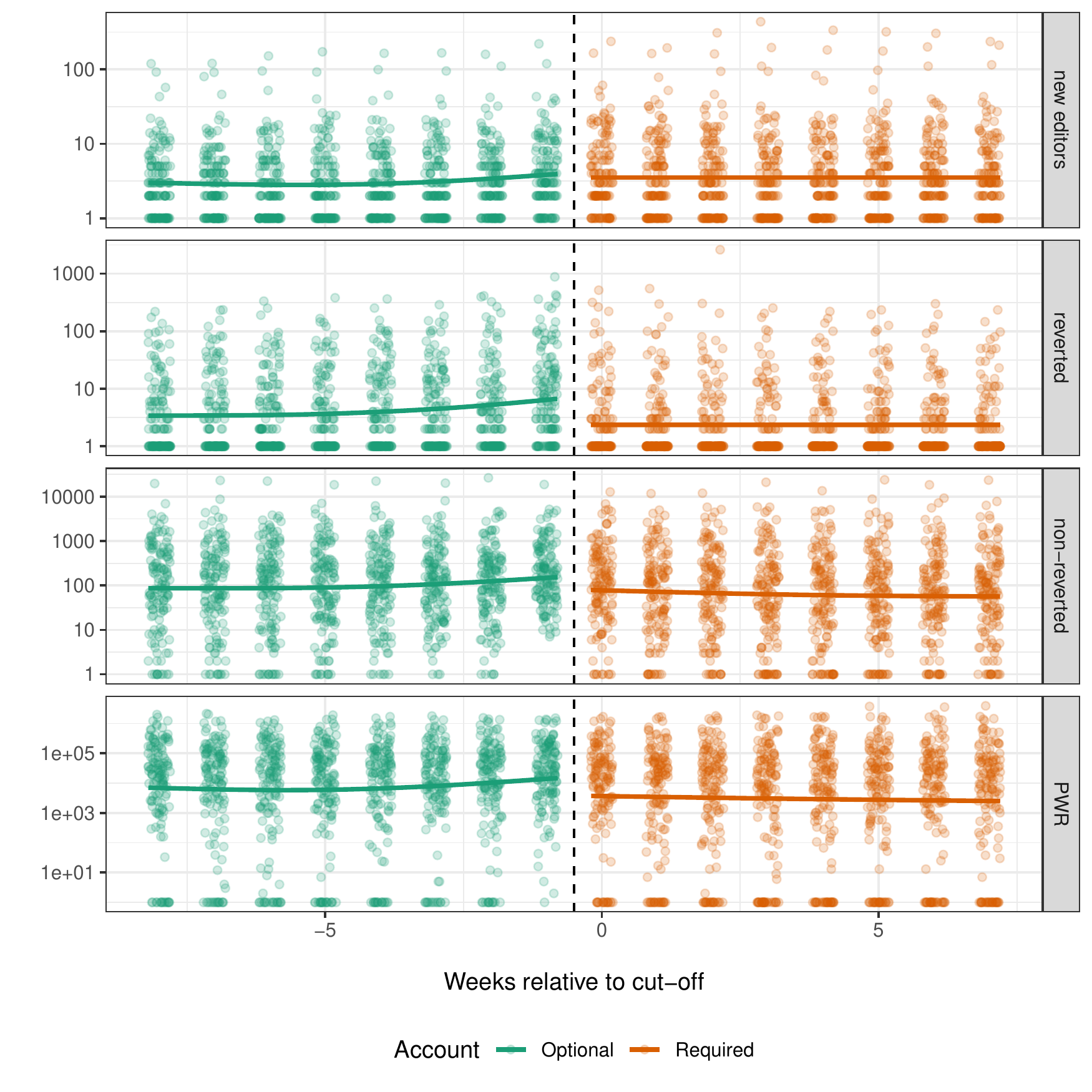} 

\end{knitrout}
\caption{Plot of all dependent variable values for all wikis over all weeks in the dataset. Points are jittered horizontally within weeks, one outlier is dropped from the reverted edits plot (the second panel from the top), and Y-axes are all log-transformed to facilitate interpretation. Central trend lines before and after the cut-off generated with a generalized additive model (GAM) smoother.}
\label{fig:dvs}
\end{figure}

Using the XML database dumps, we construct an analytic dataset where the unit of analysis for the study is the ``wiki week.'' The dataset includes 2,171 observations from a total of 136 wikis.
We use four measures of our outcome variables---each of which varies within wiki over time.
Tables of descriptive statistics summarizing the size and activity of the wikis in the sample at the point of the intervention as well as the four dependent variables in our analytic dataset appear in our online supplement. Figure \ref{fig:dvs} presents scatterplots with GAM-smoothed trend lines before and after the cut-off for all four dependent variables across the full 16-week analytic window. In each panel, we observe a large amount of variation as well as a series of discontinuous shifts at the cut-off point.

For our first dependent variable (\emph{H1}), we count the number of user accounts on each wiki that have made a contribution for the first time in the history of that wiki during each week (\emph{new accounts}). Because the XML databases only include data on the contributions to each wiki, this measure will not reflect accounts that were created but that never went on to make a contribution. If a user had created an account previously, but never made a contribution to the wiki in question, the account will be marked as new during the week it makes its first contribution.

Our second hypothesis (\emph{H2}) looks at low quality edits which we measure by counting the number of edits that are subsequently removed (\emph{reverted}) in their entirety. We adopt the simplest and most widely used approach to detecting reverts by focusing on what are called ``identity reverts'' \citep{priedhorsky_creating_2007, halfaker_dont_2011, piskorski_testing_2017}.\footnote{We detect reverts using the \textit{python-mwreverts} software library published by Aaron Halfaker at the Wikimedia Foundation. See: \url{https://github.com/mediawiki-utilities/python-mwreverts}} Using the identity revert approach, a contribution (\emph{A}) is considered reverted if and only if a second user's contribution (\emph{B}) returns the page to a state that is identical to its state before edit \emph{A}.
The affordances of MediaWiki make it easy for any user user who can edit (with or without an account) to create identity reverts and the software has affordances that allow administrators and users with special privileges to do so even more easily.
Because a large proportion of reverted contributions involve vandalism and other intentionally disruptive contributions, identity reverts provide an efficient and conservative measure of damage \citep{flock_revisiting_2012}. 
In Wikia, reverts are relatively rare \citep{teblunthuis_revisiting_2018},
and only 1 out of every 31 edits in our dataset is reverted. 10 wikis in our sample experience no reverts during our analytic window.

We construct two measures to test the effect of the intervention on higher quality contributions (\emph{H3}). First, we count the number of contributions made to a wiki in a given week that are not subsequently reverted (\emph{non-reverted}). Because the effort that goes into an edit can vary greatly, a single non-reverted contribution might reflect a large amount of new high quality content, the removal of some content added previously by another user, or the fix of a simple typo.

Our second more conservative measure of quality addresses the limits of the first by relying on the collective judgment of a wiki's contributors. Computer scientists \citet{adler_content-driven_2007} developed an approach that defines higher quality wiki content as material that persists through more subsequent contributions by others.
The most widespread operationalization of this measure is \emph{persistent word revisions (PWR)}. The PWR for a given contribution is the sum of its tokens (i.e., words and formatting elements) multiplied by the number of subsequent versions of the page (within some window) in which each token persists.\footnote{A detailed description of the method and a bibliography of papers that developed and employ the technique is available at: \url{https://meta.wikimedia.org/wiki/Research:Content_persistence} (archived at: \url{https://perma.cc/JB6W-5UB5}).  \citet{flock_toktrack_2017} describe some limitations of the approach we use.}
To construct this measure, we use the \textit{python-mwpersistence} software library published by the Wikimedia Foundation.\footnote{\url{https://github.com/mediawiki-utilities/python-mwpersistence}}
Because highly edited pages have more revisions which create more opportunities for token revisions, we follow \citet{adler_content-driven_2007} and sum the number of persistent tokens over a fixed 7 revision window.

We select the dependent variables described above because we believe they provide the best available indicators of overall wiki community health and well-being. Wikis that attract new contributors and generate more non-reverted contributions or contributions that last longer are thriving wikis. These measures are also consistent with the theoretical predictions we develop above as well as prior empirical approaches to modeling attempts to build public information goods in online communities \citep{cheshire_social_2008, halfaker_dont_2011, schweik_internet_2012, zhang_group_2011}. Differentiating between contributions of low (\emph{H2}) and high (\emph{H3}) quality allows us to consider the impact of the intervention in a more holistic fashion than much of this earlier work.

Our key independent variable is a measure of time in weeks centered on the date that each wiki blocked anonymous contributions (\emph{week}). We also include a dichotomous variable (\emph{acct\_req}) that captures whether a wiki had blocked editing by users without accounts during each week (i.e., when $\mathit{week}\geq0$). This point occurred at different moments in calendar time because wikis instituted the requirement individually. 

\subsection{Analytic Strategy and Models}

We model the effect of wikis' transition to require user accounts with a technique inspired by \emph{regression discontinuity design} (RDD).\footnote{\citet{murnane_methods_2011} provide an excellent introduction to RDD studies.} An RDD seeks to identify causal effects by comparing outcomes before and after a threshold value marking an exogenous change in a ``forcing variable'' \citep{lee_regression_2010}.
Valid inference relies on an assumption that outcomes are equal in expectation immediately around the threshold. Of course, the dependent variables (number of newcomers, reverted edits, etc.) might tend to increase or decrease around the threshold. RDD models this underlying ``secular'' trend over the forcing variable to create a more precise estimate of the change attributable to the discontinuity.

In this case, we adapt RDD methods to a within-subjects analysis with repeated observations (panel data) for every subject (wiki) capturing each subject's individual secular trend.
The forcing variable is a relative measure of time ($\mathit{week}$) centered at the discontinuity. Our estimand ($\tau$) is the average effect of the intervention on the communities that underwent the change.
The parameter estimate $\hat\tau$ associated with $\mathit{acct\_req}$ captures any discontinuous shift in the outcomes at the point the wikis required accounts.

Our panel RDD approach shares some features with interrupted time series (ITS) and provides insight into the instantaneous effects of requiring accounts on the treated communities. We adopt this ``within subjects'' panel RDD approach for two reasons. First, it lets us compare the treated wikis to themselves immediately before and after the treatment in the absence of control cases.  We do so because true control cases do not exist given the quasi-experimental nature of the change. Although matching on observables is often used to construct control cases in such situations, our preliminary evaluations and experience of Wikia led us to believe that wikis which implemented the account requirement likely differed in unobserved ways from wikis that did not. Second, we estimate the instantaneous effects because differences in outcomes farther from the discontinuity are more likely due to sources of heterogeneity unrelated to the account requirement. We discuss these choices and assumptions as well as alternative approaches at greater length in the online supplement.

Our models each include 408 fixed effects control variables. First, we include a vector of dummy variables \textbf{\emph{wiki}} that constitute wiki-level fixed effects. To model trends within each wiki, we interact the \textbf{\emph{wiki}} vector with both $\mathit{week}$ and $\mathit{week}^2$.\footnote{Because of visual and statistical evidence of curvilinear secular trends within many of our wikis, we use a quadratic specification.
Although a cubic model had improved goodness of fit, there was no substantive change in $\hat{\tau}$, so we report the more parsimonious model.}
Fixed effects in repeated measure models are typically considered preferable to random effects because their inclusion ensures that estimates of $\hat\tau$ reflect only within group variation \citep{angrist_mostly_2008, murnane_methods_2011}.
In this way, our three vectors of fixed effects capture all observed and unobserved characteristics of wikis that have a consistent effect across all weeks (such as each project's start date or initial audience size) as well as a separate quadratic trend for each wiki (such as growth or decline in the popularity of a topic covered by a wiki).
As our hypotheses focus exclusively on the estimate of $\hat\tau$, the fixed effects serve exclusively as controls and we do not report the parameter estimates for them.
The models thus take the form: 

\vspace{-2em}
 
\begin{equation}
\mathit{Y}= \tau \mathit{acct\_req} + \bm{\beta} \bm{\mathit{wiki}} + \bm{\beta} (\bm{\mathit{wiki}} \times \mathit{week}) + \bm{\beta} ( \bm{\mathit{wiki}} \times \mathit{week}^2 ) + \varepsilon
\end{equation}

\noindent We set $\mathit{Y}$ equal to each outcome measure in turn: \emph{new users} (M1), \emph{reverted} (M2), \emph{non-reverted} (M3a), and \emph{PWR} (M3b). To evaluate the effect of the intervention on contributions from experienced community members (\emph{H4}), we re-estimate models 2-4 restricting the sample to editors who had made at least one contribution under a registered account prior to the cut-off. 

Since all the dependent variables are skewed counts, we use negative binomial regression.\footnote{Given the relatively small amount of variation in \emph{reverts} in our sample, we report a logistic regression model of whether wikis experience any reverted edits in a section titled ``Logistic Specification for Reverts'' in our online supplement. The results suggest a similar takeaway.} We calculate heteroskedasticity-robust standard errors using the $HC_1$ estimator \citep{mackinnon_heteroskedasticity-consistent_1985} and follow \citepos{angrist_mostly_2008} suggestion to report the larger of robust or conventional standard errors.
Finally, we conduct a supplementary analysis that decomposes $\hat{\tau}$ across each wiki in our sample. To do this, we re-estimate our models adding a term interacting \textbf{\emph{wiki}} with $\mathit{acct\_req}$ to recover individual wiki-level estimates for $\hat{\tau}$.
Although some wikis in our analysis had contributions from unregistered editors after the cut-off,  we estimate the impact of the intervention irrespective of compliance. Our results thus correspond to an \emph{Intent-to-treat (ITT)} estimate \citep{murnane_methods_2011} in that we measure the impact of a change even though it may not have been perfectly effective in every case.

\section{Results}

We find substantial discontinuous shifts in each of our dependent variables when wikis began requiring accounts: an increase in the number of new editors and decreases in the numbers of reverted edits, non-reverted edits, and PWR. These outcomes support hypotheses \emph{H1}-\emph{H3}. We also find similar decreases among editors who contributed with accounts prior to the intervention, supporting hypothesis \emph{H4}.

\begin{figure}
\begin{knitrout}
\definecolor{shadecolor}{rgb}{0.969, 0.969, 0.969}\color{fgcolor}
\includegraphics[width=\maxwidth]{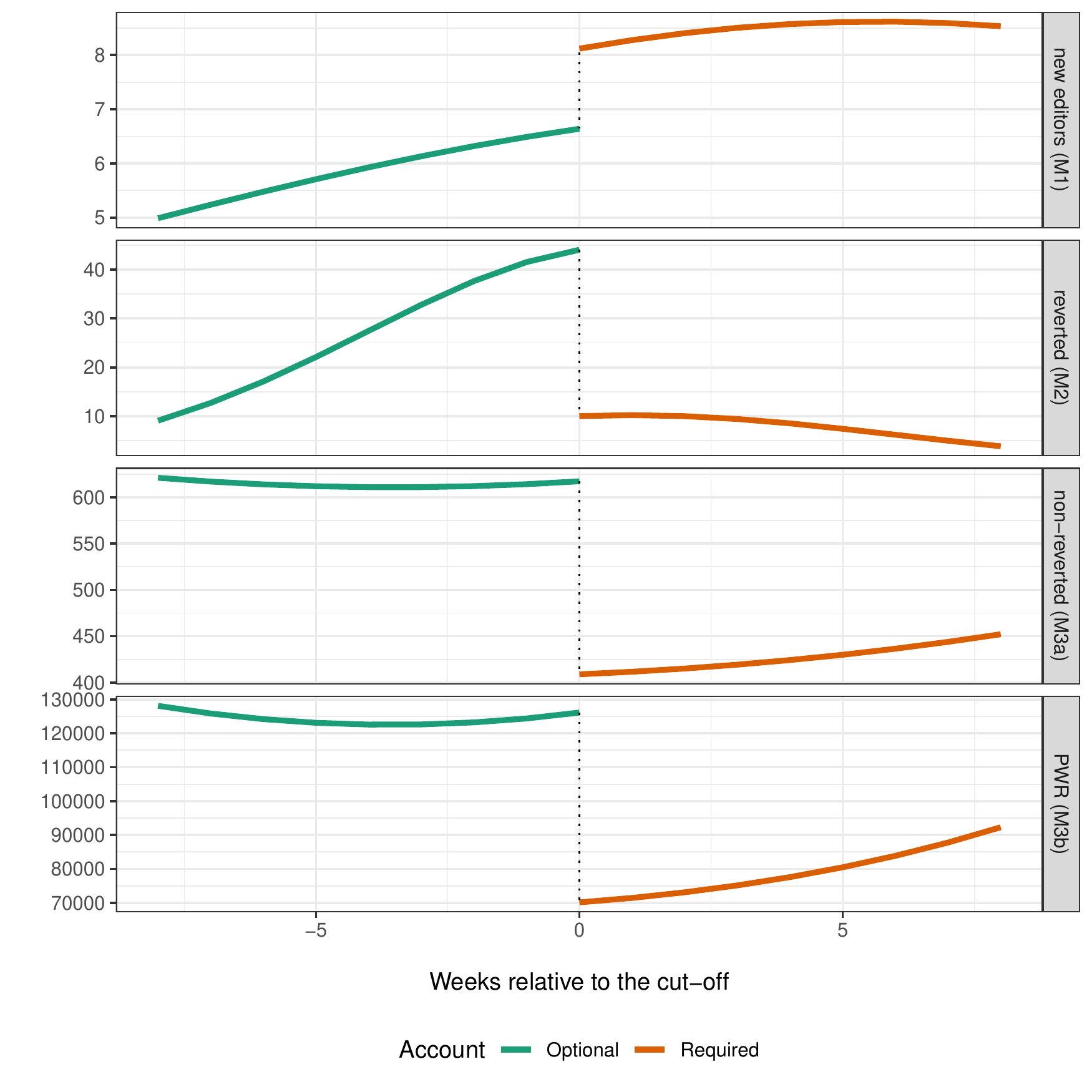} 

\end{knitrout}

\caption{Visualizations of our model predicted values for prototypical wikis for all four outcomes. To represent the effect of an ``average'' wiki in our sample, trend lines reflect a fitted regression model of window weeks and the cut-off on the predicted values from each of our models reported in Table \ref{tab:reg}.
  We do not plot predicted values corresponding to \emph{H4} because the estimates align so closely with those for \emph{H3} (\emph{M3a} and \emph{M3b}).}
\label{fig:protoplots}
\end{figure}

Table \ref{tab:reg} summarizes the results of \emph{H1}-\emph{H3} and includes regression models fit using the full sample. We report the estimated parameter ($\hat{\tau}$) associated with \emph{acct\_req} from each model. Because these are non-linear models, interpreting the coefficients directly can be challenging. In order to facilitate interpretation, we also present the model-predicted values for prototypical wikis in Figure \ref{fig:protoplots}.\footnote{Figure \ref{fig:protoplots} summarizes an ``average'' trend by plotting predicted values from a negative binomial model that regresses the fitted values generated by the original models on a quadratic specification of week and the indicator for the cut-off.}

\begin{table}
  \centering
\begin{adjustbox}{center}
\begin{tabular}{l c c c c }
\hline
 & \emph{new editors} (M1) & \emph{reverted} (M2) & \emph{non-reverted} (M3a) & \emph{PWR} (M3b) \\
\hline
acct\_req                                                       & $0.201^{***}$      & $-1.477^{***}$     & $-0.412^{***}$   & $-0.586^{***}$    \\
                                                                & $(0.057)$          & $(0.162)$          & $(0.092)$        & $(0.118)$         \\
\hline
Deviance                                                        & 2157.498           & 1612.692           & 2543.870         & 2599.352          \\
Num. obs.                                                       & 2171               & 2171               & 2171             & 2171              \\
\hline
\multicolumn{5}{l}{\scriptsize{$^{***}p<0.001$, $^{**}p<0.01$, $^*p<0.05$}}
\end{tabular}

\end{adjustbox}
\caption{Summary of regression models. We omit estimates for our 408 control variables in the form of fixed effects for \textbf{\textit{wiki}} and its interaction with \textit{week} and \textit{week}\textsuperscript{2}.}
\label{tab:reg}
\end{table}

Model M1 supports \emph{H1} that requiring accounts leads to a higher number of new user accounts. We estimate that the number of accounts making their first contribution rose from 6.64 to 8.12 new users per week in a prototypical community at the point that they required accounts ($\hat{\tau}=0.201, \mathrm{SE}=0.057, p<0.001$). This change constitutes a 22\% increase. However, the standard errors are relatively large and the 95\% confidence interval for the parameter estimate corresponds to an increase of between 0.62 and 2.44 new accounts (9\% and 37\% of the total in a prototypical wiki).
Over the course of eight weeks following the change, we predict a total number of new accounts that is 
122\% what we would have predicted in the absence of the shift.

Model M2 provides strong support for \emph{H2}, that requiring accounts drives down levels of low quality edits. M2 estimates a decline from 44.04 to 10.03 reverted edits per week (77\%) in a prototypical community ($\hat{\tau}=-1.477, \mathrm{SE}=0.162, p<0.001$). Our 95\% confidence interval corresponds to a decrease between 36.71 and 30.25 reverted contributions (83\% and 69\% of the total in a prototypical community). For less active wikis in our sample this might involve a near-complete elimination of reverted edits.
Over the course of eight weeks, we predict a total number of reverts that is 23\% what we would predict with unregistered contribution permitted.

In Models M3a and M3b we find that the block also caused a decrease in quality edits (supporting \emph{H3}). Using our first measure of quality, M3a shows that the cut-off led to a decrease from 617 to 409 non-reverted contributions each week (34\%) in a prototypical community ($\hat{\tau}=-0.412, \mathrm{SE}=0.092, p<0.001$). Our 95\% confidence interval for this parameter estimate encompasses between 276 and 128 contributions per week (i.e., decreases of 45\% and 21\% of the total).
Over eight weeks, we predict a total number of non-reverted edits that is 66\% what we would have predicted otherwise.
The estimates in M3b are very similar to those in M3a. We estimate that the intervention led to a decrease of 44\% from 126,000 to 70,200 PWR per week at the cut-off  in a prototypical community ($\hat{\tau}=-0.586, \mathrm{SE}=0.118, p<0.001$).
with 95\% confidence intervals between 70,400 and 37,900 word revisions per week (i.e., decreases of 56\% and 30\% of the total).
Over eight weeks, we predict a total number of PWR that is 56\% the predicted total in the absence of the shift.

\begin{table}
    \centering
\begin{tabular}{l c c c }
\hline
 & \emph{reverted} (M2) & \emph{non-reverted} (M3a) & \emph{PWR} (M3b) \\
\hline
acct\_req                                                       & $-0.794^{***}$     & $-0.455^{***}$   & $-0.473^{***}$    \\
                                                                & $(0.198)$          & $(0.093)$        & $(0.121)$         \\
\hline
Deviance                                                        & 1168.152           & 2519.045         & 2508.188          \\
Num. obs.                                                       & 2164               & 2164             & 2164              \\
\hline
\multicolumn{4}{l}{\scriptsize{$^{***}p<0.001$, $^{**}p<0.01$, $^*p<0.05$}}
\end{tabular}

    \caption{Summary of regression models estimated on sample restricted only to editors whose accounts were created prior to the cut-off. Estimates for our 408 fixed effects for \textbf{\textit{wiki}} and its interaction with \textit{week} and \textit{week}\textsuperscript{2}.}
\label{tab:reg.noprecutoff}
\end{table}

The model results on the restricted sample of editors with accounts prior to the cut-off (corresponding to \emph{H4}) is shown in Table \ref{tab:reg.noprecutoff}.
M1 is not included because all accounts among editors active before the cut-off must have been created before.
The results for Models M2, M3a, and M3b on the restricted sample suggest very similar patterns to those from the full sample.
M2 finds that the number of reverted edits among experienced editors undergoes a discontinuous decrease after the intervention that is smaller in magnitude than in the full sample. Models 3a and 3b also show a discontinuous decrease in the number of non-reverted edits and PWR respectively. The magnitudes of each estimate are very similar to those reported in the main analysis.
This follows the pattern observed in both \citet{gorbatai_paradox_2014} and \citet{kane_emergent_2014}, where contributions from unregistered editors stimulate contributions from existing editors.

In order to understand the heterogeneity of the effect among wikis, we fit a series of models incorporating an interaction between \textbf{\emph{wiki}} and $\mathit{acct\_req}$. These models generate a separate point estimate of $\hat{\tau}$ for each wiki. The results, described in our online supplement, show substantial heterogeneity. For M1, we find that \emph{more} wikis show a decrease in new editors (73) than an increase (60) suggesting that our conclusion in terms of H1 may be driven by large increases among a subset of wikis. 
In the case of M2, we estimate that more than 3 wikis experience a decrease in reverts for each that experiences an increase (102 wikis with a decrease versus 33 wikis with an increase). The results for M3a and M3b each suggest that a little over 60\% of wikis experienced a decrease (respectively, 86 and 83 with negative estimates, 50 and 52 with positive estimates, and 1 with an estimate of zero in M3b).
Overall, this suggests that aggregate estimates accurately characterize the trends among wikis while demonstrating that heterogeneous effects also occurred.

\section{Threats to Validity}

Several threats to the validity of our findings stem from challenges common to RDDs and the RDD-inspired approach we adopt \citep{lee_regression_2010, murnane_methods_2011}. Below, we briefly discuss the most salient threats and, when possible, the sensitivity analyses used to assess them. We also revisit the assumptions that underpin the validity and generalizability of our findings. In our online supplement, we present additional analysis, technical threats, and the full results of all sensitivity analyses and alternative model specifications. 

Administrators' knowledge of the intervention poses arguably the most critical threat to our assumption of the exogeneity of the precise timing of the intervention in relation to the dependent variables.
This threat, a specific case of what is called ``crossover'' in quasi-experimental methods, arises from the fact that some community administrators would have known of the intervention ahead of time and could have manipulated their pre/post-treatment behavior.
Although it is impossible to prove the absence of crossover, we test the sensitivity of the findings to the inclusion of the administrators in the analysis by dropping edits from administrators.
The results are similar but the effects are slightly larger in magnitude.

In other sensitivity analyses, the effects of the intervention on low and high quality edits are robust to alternative specifications. Under all approaches explored in the supplement, our main analysis estimates for M2, M3a, and M3b retain the same sign while differences in the magnitude of the point estimates under these alternative specifications are smaller than the original standard errors. In contrast, the effects on new account creation (M1) are more fragile. Under several specifications, the point estimate shrinks substantially or becomes indistinguishable from zero. We also find that the increase in \textit{new accounts} estimated in the main analysis is not robust to dropping the largest wikis from the analysis (see the subsection on ``Influential Cases'' in the online supplement). The estimates of \textit{acc\_req} in these models are consistently positive but smaller in magnitude and are not statistically significant in models that drop the largest 5\% and 10\% of wikis. Although we can imagine that the effect of requiring accounts on new editors might vary with project size or other factors, we know of no theoretical reasons to anticipate this. It is also possible that the fragility of the estimate is an artifact of fact that the effect on new editors is relatively small and noisy and that much of the estimated effect is concentrated in large wikis. As a result of these tests, the results of M1 in Table \ref{tab:reg} should be interpreted with care.

Our panel RDD approach rests on the assumption that individual wikis were ``equal in expectation'' to themselves immediately before and after the point of the intervention. As discussed earlier, we are concerned that administrators of the treated wikis played a role in the decision to require accounts.  We have tried to evaluate this assumption through both qualitative investigation (i.e., conversations with wiki administrators and manual review of on-wiki discussions by participants) as well as empirical sensitivity analysis described above and reported in the online supplement. Although we find no likely causes of bias in this regard, we cannot eliminate the possibility of this threat.
In the absence of an empirical matching strategy or credible control group with which to construct alternative baselines for estimating treatment effects, we do not claim that our results generalize to untreated wikis. Additional assumptions and further analysis would be necessary to extend our analysis in this way.\footnote{See our online supplement for further discussion of this issue.}

Another threat is that the imposition of an account requirement may have affected our measures in M2 and M3a. 
Before the account requirement, unregistered editors may have participated in the work of reverting low-quality edits in ways that were made more difficult by the requirement to create accounts. Although editors with accounts might have reverted the same edits themselves---perhaps after a longer period of time---it is possible that some reverts of low-quality edits might never have occurred after accounts were required. In this scenario, our estimate of a decrease in \emph{reverted} might over-estimate the change in low-quality contributions and our estimate of a decrease in \emph{non-reverted} might be under-estimate the decrease in high quality edits. Although an empirical investigation suggests that only 1,450 of the 19,236 reverting edits (7.5\%) made in the 8 week period before wikis required accounts were made by users without accounts, we cannot know whether this threat affects our estimates. 
We can gain comfort from the fact that, to the extent that this threat obtains, it means that our estimates of collateral damage caused by requiring accounts and the tradeoff that communities face will both be rendered more conservative.
 
Tests of the external validity of our estimates lie beyond the scope of the current paper. Our central finding that requiring accounts causes a decline in both low quality and high quality contributions resonates with related work in the contexts of citizen science \citep{jay_sign_2016} and discussion forums \citep{bernstein_4chan_2011}. In particular, \citet{jay_sign_2016} estimate an increase of over 60\% in the number classifications to a citizen science platform when accounts are not required. The Wikimedia Foundation has also released an unpublished study with comparable point estimates of a $25-30$\% decrease in overall productivity when accounts were requested of would-be contributors.\footnote{\url{https://meta.wikimedia.org/wiki/Research:Asking_anonymous_editors_to_register} (archived at: \url{https://perma.cc/2H48-77GT}).} We believe that these data points provides reasons to be optimistic about the generalizability of our results.

Lastly, our estimates may not generalize to other types of account requirements. Although the design of the Wikia account creation and sign-in system resembles similar systems on a number of peer production projects, social media platforms, and online discussion communities, the system design could be adjusted to streamline the experience of account creation, emphasize the value of stable identifiers, or incorporate other elements. We anticipate that such alternative designs could alter the effects we estimate here.

\section{Discussion}

Our results support our hypotheses and suggest that stable identifiers impose barriers to entry that introduce a tradeoff for online communities. While the account requirement drastically redcuces low quality contributions and induces some individuals to create new accounts, it also decreases high quality participation enormously.
We find evidence for this tradeoff even within the editing activity of editors registered prior to the cut-off demonstrating how participation by unregistered editors stimulates activity across the board. Accumulated over many weeks, these effects would profoundly change the scope and character of participation within these projects.

The results diverge from earlier findings and theories consistent with the ``catalysts'' perspective which would have predicted stable or increased rates of high quality contributions. What explains this difference? Differences between experimental contexts \citep[e.g., ][]{yamagishi_solving_2009} and the wikis in our sample offer one potential explanation. Perhaps communities that require accounts for participation at the point they are initially constituted (like those in most experiments) experience different effects than existing communities (like those in our sample). It might also be the case that specific institutional arrangements support some types of public goods production or collective action better than others. In this sense, our findings may be interpreted as supporting the idea that particular types of community boundaries may promote or hinder particular sorts of common pool resource management \citep{cox_review_2010}. The account requirements in our study may also lead to subtle long-term effects on community culture that we do not estimate or observe due to the constraints of our measures and analytic approach.

Although our findings resonate more with the costly barriers perspective, they deviate from it as well. The decline we see among registered editors does not follow from a pure transaction costs approach which might also predict that these registered editors would contribute more new content (i.e., PWRs) following the removal of low quality participation that may have required them to focus on vandal fighting. Instead, unregistered contributions---even the low quality ones---seem to induce community activity. 
This supports earlier claims about the value of group size and peripheral participation for collective action and public goods production in online groups or crowds \citep{anthony_reputation_2009, bennett_logic_2012, gorbatai_paradox_2014, kane_emergent_2014, margolin_emotional_2018, olson_logic_1965, woolley_evidence_2010, zhang_group_2011}. Our analysis cannot unpack this finding more fully because our data does not include additional information about the individual contributors who had participated without accounts prior to the cut-off. Future work should analyze the content of unregistered contributions more deeply to understand the value they add to communities and use network analysis and related methods to explain the structural mechanisms by which unregistered contributions enhance group collaboration.

Our theory that community boundaries introduce tradeoffs by reducing both low and high quality contributions suggests some possible common ground between the catalysts and costs perspectives. Even a strong proponent of the catalysts point of view might not be surprised that an imprecise intervention like an account requirement would deter some high quality contributions. However, the lopsided empirical tradeoff we observe in our results strongly favors preserving opportunities for unregistered contribution. 
For example, in the prototypical wiki shown in Figure \ref{fig:protoplots}, our models estimate that the account requirement will deter 242 edits but that only 34 of those contributions (14\%) would have been reverted.
In order to justify requiring accounts in our empirical context, the damage prevented by eliminating any single reverted edit would need to outweigh the loss of 7 non-reverted contributions. 
Given that removing damage in peer production is typically easier than substantive content development, we find it hard to imagine how these findings could be used to support arguments in favor of requiring accounts. In that sense, our estimates will likely surprise both the administrators of the wikis in our sample who took steps to require accounts as well as ``catalysts'' theorists whose work supported the decision to do so. 

At the same time, if we focus on the results in terms of relative proportions (instead of the raw numbers of low and high quality contributions), a sufficiently large decline in low quality contributions could provide benefits that offset a smaller decline in high quality contributions. Similarly, if each individual low quality act were incredibly damaging and costly---something that does not seem to be the case in our setting---the tradeoff might be seen as worthwhile. Such scenarios extrapolates beyond the current study in a number of ways and introduces risks of generalizing outside our population.

In any case, the visible benefits of reduced vandalism or spam must be weighed against the hidden costs of reduced high quality contributions. Theories in this space need to more fully accommodate the tensions between open community boundaries and effective safeguards of contribution quality. The empirical tradeoff at the heart of our analysis suggests that requiring accounts will not benefit most communities in the long run. Instead, our findings suggest that most communities should preserve opportunities for unregistered contributions. While unidentified and anonymous participation may bring challenges around disinhibition, lack of accountability, and breakdowns of trust, our results indicate that requiring accounts erects a boundary around participation that undermines the capacity of open communities to produce valuable public goods. To fight low quality contributions, these communities might seek more precise barriers to entry or might attempt to lower the cost of dealing with undesirable forms of participation.

\section{Conclusion}

Our study contributes to communication research on public information goods, stable identifiers in computer-mediated communication, information systems, and online communities and crowds. We provide the first large-scale observational evidence of the causal effects of requiring accounts in peer production systems. The tradeoffs we theorize and observe synthesizes prior research that implied incongruous empirical predictions. Our findings reveal limitations of this earlier work and show that a specific implementation of community boundaries through imposing the use stable identifiers impacts participation in multiple ways. Overall, the account requirement decreases both low and high quality contributions to the wikis in our study. Although it deters a large portion of all low quality participation, the vast majority of deterred contributions are high quality. Most surprisingly, we observe a large decrease among contributors who hold accounts already and are not directly impacted by the design change.

The findings illustrate how the empirical tradeoffs play out across diverse communities. We conclude that the results support the preservation of opportunities for unregistered, low cost forms of contribution to public information goods online. However, these implications depend on contextual factors such as the costs of dealing with low quality participation. For example, automated tools to filter spam can scale well with minimal overhead. Large projects such as the English language Wikipedia have anti-vandalism bots that effectively minimize the cost of removing many low quality contributions. As a result, Wikipedia can afford to enable unregistered editing and draw benefits from good-faith contributors who would not participate otherwise. This illustrates how the nature of contribution costs likely also matter more than prior work had suggested. Completing a CAPTCHA is not the same as creating an (even pseudonymous) account on a site. Both the content and framing of interfaces shape how newcomers experience the boundaries of a community organization and interpret the significance of participation. 

The results of this study suggest several directions for future research. A field experiment could support valid comparisons of communities with and without account requirements over longer time periods. Subgroup analyses might decompose the outcomes we observe here to identify particular types of wiki communities that experienced the effects of the intervention to varying degrees. Other approaches can support deeper investigation of the mechanisms by which unregistered contributions support overall group activity. Community managers and systems designers might also investigate alternative boundary management or quality control systems as means of promoting the sort of prosocial cooperation and collective action identified in previous research. For example, prior work has described system designs that support privacy as well as safety in the moderation of user-generated content \citep{frey_can_2017}. 
Our findings speak to the importance of continuing to test and refine both the theories and empirical approaches in these domains.

\section{Data and Supplementary Material}

We have made available the full data and code necessary to reproduce this analysis and this article and a comprehensive online supplement with additional technical details and the results of robustness checks in the Harvard Dataverse (\url{https://doi.org/10.7910/DVN/CLSFKX}).

\section{Acknowledgments}

The authors gratefully acknowledge audiences at ETH Zurich Center for Law \& Economics, UC Berkeley MORS at the Haas School of Business, the ANN-SONIC Conference on Peer Production Networks, the Information Society Project at Yale Law School, the Berkman Klein Center for Internet \& Society at Harvard University, and the BYOR Workshop at Northwestern for feedback on earlier versions of this work. Computing resources for the project were supported and managed through the Hyak shared scalable compute cluster for research at the University of Washington. Yochai Benkler inspired the project and Danny Horn showed us the way to the data. Our collaboration benefited from time at both the Helen Riaboff Whiteley Center at the University of Washington Friday Harbor Labs and the Center for Advanced Study in the Behavioral Sciences at Stanford University. Support for this research was provided by Northwestern University, the University of Washington, and the United States National Science Foundation (IIS-1617129, IIS-1617468, and CNS-1703049).

\printbibliography[title = {References}, heading=secbib]

\end{document}